\documentstyle[aps,prl,twocolumn,epsf,graphics,psfig]{revtex}

\begin{document}
\draft

 \newcommand{\mytitle}[1]{

 \twocolumn[\hsize\textwidth\columnwidth\hsize

 \csname@twocolumnfalse\endcsname #1 \vspace{1mm}]}


\title{Boltzmann  theory of engineered anisotropic magnetoresistance in (Ga,Mn)As} 
\author{T. Jungwirth$^{1,2}$\cite{ramin2}, M. Abolfath$^{3}$\cite{ramin2}, Jairo Sinova$^{2}$, 
J. Ku\v{c}era$^{1}$, and A.H. MacDonald$^{2}$}
\address{$^{1}$Institute of Physics ASCR, Cukrovarnick\'a 10,
162 53 Praha 6, Czech Republic\\}
\address{$^{2}$Department of Physics,
The University of Texas at Austin, Austin, TX 78712 \\}
\address{$^3$Department of Physics and Astronomy, University of Oklahoma, 
Norman, OK 73019-0225}
\date{\today}
\maketitle
\begin{abstract}
We report on a theoretical study of dc transport coefficients in
(Ga,Mn)As diluted magnetic semiconductor ferromagnets that accounts for 
quasiparticle scattering from ionized Mn$^{2+}$ acceptors with a local
moment $S=5/2$ and from non-magnetic compensating
defects.  In metallic samples Boltzmann transport theory with Golden rule scattering rates 
accounts for the principle trends of the measured 
difference between 
resistances for magnetizations parallel and perpendicular to the 
current.  We predict that the sign and magnitude of the 
anisotropic magnetoresistance can be changed by strain engineering or by altering 
chemical composition.   
\end{abstract}


In most (III,V) semiconductors, Mn$^{2+}$ 
substitution on a cation (column III element) site 
introduces an $S=5/2$ local moment and a valence band hole\cite{szczytkoprb99-01}.
Mn$_x$III$_{1-x}$V diluted magnetic semiconductors
\cite{ohnosci98,ohnojmmm99,potashnik0204250,gallagher0205517}
(DMSs) are ferromagnetic and metallic for Mn fractions larger than $x\sim 1\%$. 
Many magnetic properties of the most robustly ferromagnetic samples,  those that
have $\sim 5\%$ or more Mn and are annealed to reduce the density of compensating defects, 
appear to be adequately explained by virtual crystal approximation models in
which disorder is ignored 
\cite{dietlsci00,dietlprb01,abolfathprb01,bookchapter}.
For example in bulk DMSs, this approach can account for ferromagnetic
critical temperatures $\sim$100~K \cite{dietlprb01,jungwirthprb02}, 
the correlation between magneto-crystalline anisotropy and substrate lattice constant
\cite{dietlprb01,abolfathprb01}, 
the size and sign of the anomalous Hall effect \cite{jungwirthprl02}, 
and several optical properties \cite{dietlprb01}.
It has also been used to describe properties of DMS heterostructures 
\cite{boukariprl02,ramin,leesst02} 
for which the simplifications afforded by neglect of disorder are particularly helpful.

In this letter we investigate theoretically the $T=0$ dc-transport coefficients of
(Ga,Mn)As ferromagnetic semiconductors. We find that relaxation-time-approximation solutions
of the Boltzmann equation provide anisotropic magnetoresistance ($AMR$)
estimates that are
in good agreement with experiments \cite{baxter0202508,gallagher}. 
Our results suggests that transport properties of these
metallic ferromagnets can be understood within a conventional framework in which disorder
is treated as a weak perturbation. 
We find that the conductivity 
varies by several percent when the magnetic order parameter is reoriented by a weak
magnetic field, and 
predict that the magnitude and sense of this change depends on the chemical composition
and on the substrate on which the thin film DMS ferromagnet
is epitaxially grown.  This spontaneous magnetoresistance anisotropy is the transport analog of
magneto-crystalline anisotropy \cite{dietlprb01,abolfathprb01} which has approximately 
the same size relative to the total condensation energy of the ordered state.   
All results presented in this paper are for the (Ga,Mn)As DMS's.
A large database that details our predictions for the $AMR$ 
of many other host semiconductors over a wide range of compositions and strains
is available on the internet 
\cite{msweb}.

We consider a microscopic Hamiltonian in which valence band holes interact with randomly
located spins of substitutional Mn$^{2+}$ impurities via exchange interactions, 
and with randomly located ionized defects and each other via Coulomb interactions.  
Focusing on $T=0$, we assume that the Mn spins are fully 
aligned in the ferromagnetic ground state.
In the virtual crystal approximation, the interactions are replaced by their 
spatial averages, 
so that the Coulomb interaction vanishes and hole quasiparticles 
interact with a spatially constant Zeeman field. 
The unperturbed Hamiltonian for the holes then reads
$H_0=H_L+J_{pd}N_{Mn^{2+}} S \hat \Omega \cdot\vec s\;$,
where $H_L$ is the six-band Kohn-Luttinger Hamiltonian\cite{abolfathprb01}, 
$\hat \Omega$ is the Mn local moment orientation, $J_{pd}=55$~meV nm$^3$ \cite{ohnojmmm99}
is the local-moment -- valence-band-hole kinetic-exchange coupling constant, 
$N_{Mn^{2+}}$ is  the
density of ordered Mn local moments, 
and $\vec s$ is the envelope-function hole spin operator\cite{abolfathprb01}. 
We use the relaxation-time-approximation solution to the semiclassical Boltzmann equation
to estimate the dc conductivity tensor: 
\begin{eqnarray}
\sigma_{\alpha\beta}&=&\frac{e^2}{\hbar V}\sum_{n,k}
(\hbar\Gamma_{n,\vec k})^{-1}
\frac{\partial E_{n,\vec k}}{\partial k_{\alpha}}
\frac{\partial E_{n,\vec k}}{\partial k_{\beta}}
\delta(E_F-E_{n,\vec k}) \; ,
\label{sigma}
\end{eqnarray}
where 
$\Gamma_{n,\vec k}$ 
is the quasiparticle elastic scattering rate, 
$n$ and $\vec k$ are the band and wavevector
indices of the valence band Bloch states of the unperturbed system, and 
$E_{n,\vec k}$ are the spin-split band energies of the ferromagnetic state. 
In Eq.~(\ref{sigma}) 
we have omitted the asymmetric terms in the off-diagonal elements of
$\sigma_{\alpha\beta}$ that contribute to the anomalous
Hall conductivity, discussed in detail elsewhere \cite{jungwirthprl02}. 
The symmetric off-diagonal elements, described
by Eq.~(\ref{sigma}), vanish when the magnetization
is aligned along one of the cube edges of the host lattice.

In our model, itinerant holes are scattered on substitutional 
Mn$^{2+}$ impurities by a Thomas-Fermi 
screened Coulomb potential and by a magnetic-kinetic-exchange potential.
For majority-spin holes both potentials are attractive while for 
minority-spin holes the magnetic potential becomes repulsive. We
estimate the transport weighted scattering
rate from Mn$^{2+}$ impurities using Fermi's golden rule:
\begin{eqnarray}
\Gamma^{Mn^{2+}}_{n,\vec k}&=&\frac{2\pi}{\hbar} N_{Mn^{2+}}\sum_{n^{\prime}}
\int\frac{d\vec k^{\prime}}{(2\pi)^3} 
|M_{n,n^{\prime}}^{\vec k,\vec k ^{\prime}}|^2
\nonumber \\ &\times &\delta(E_{n,\vec k}
-E_{n^{\prime}\vec k ^{\prime}})
(1-\cos \theta_{\vec k, \vec k ^{\prime}})\; ,
\label{gamma}
\end{eqnarray}
where the scattering matrix element was approximated by the following expression,  
\begin{eqnarray}
M_{n,n^{\prime}}^{\vec k,\vec k ^{\prime}}&=&
J_{pd}S
\langle z_{n \vec k}|\hat \Omega\cdot \vec s|
z_{n^{\prime}\vec k ^{\prime}}\rangle\nonumber \\
&-&
\frac{e^2}{\epsilon_{host}\epsilon_0(|\vec k -\vec k ^{\prime}|^2
+q_{TF}^2)}\langle z_{n \vec k}|
z_{n^{\prime}\vec k ^{\prime}}\rangle .
\label{mnelement}
\end{eqnarray}
Here $\epsilon_{host}$ is the host semiconductor dielectric constant,
$|z_{n \vec k}\rangle$ is a six-component envelope-function eigenspinor of the Hamiltonian $H_0$, 
and the Thomas-Fermi screening wavevector was approximated by the parabolic band expression, 
$q_{TF}=\sqrt{3 e^2p/(2 \epsilon_{host}\epsilon_0E_F)}$, where $p$ is the itinerant hole density
and $E_F$ is the Fermi energy. 

Recent experiments have established that magnetic and
transport properties of (III,Mn)V DMS 
ferromagnets are  sensitive to post-growth 
annealing protocols
\cite{potashnik0204250,gallagher0205517,baxter0202508,yuprb02},
and that this sensitivity is associated with changes in the density of 
defects that compensate the Mn$^{2+}$ acceptors.  Our model recognizes that the 
transport properties of these materials are not determined solely by the 
scattering from substitutional
Mn$^{2+}$ impurities and allows explicitly for scattering from
compensating defects. 
We assume that compensation can occur due to the presence
of As-antisite defects (common in low-temperature MBE grown GaAs hosts)
or due to Mn interstitials. 
As-antisite defects are non-magnetic and contribute only $Z=2$ Coulomb scattering. 
Mn interstitials, when they are present\cite{yuprb02,gallagher}, 
are unlikely to be magnetically
ordered and can also be modeled as $Z=2$ donors \cite{masek0201131}.
Overall charge neutrality implies that the density of holes is 
$p=N_{Mn^{2+}}-2N_{c}$, where $N_{c}$ is the density of  
compensating impurities.

Rough estimates for the $T=0$ quasiparticle scattering rates 
can be obtained from parabolic-band-approximation 
expressions for majority heavy-hole states assuming Mn$^{2+}$ 
kinetic-exchange scattering only,  
$\Gamma_{pd}=(N_{Mn^{2+}}) J_{pd}^2 S^2 m^{\ast} \sqrt{2m^{\ast}E_F}/(4 \pi \hbar^4)$,
or Mn$^{2+}$ and As-antisite
Coulomb scattering  only which leads to scattering rate $\Gamma_{C}$ given by the
Brooks-Herring formula \cite{brooks}. 
(We show below that most of the current
in a (Ga,Mn)As ferromagnet is carried by majority spin heavy-holes.) 
Taking the  heavy-hole effective mass 
$m^{\ast}=0.5m_e$, 
$p=0.4$~nm$^{-3}$ and Mn fraction $x=5$\%, we obtain
$\hbar\Gamma_{pd}\sim 20$~meV and $\hbar\Gamma_{C}\sim 150$~meV.
Our full numerical six-band calculations 
are consistent with these estimates, and in particular predict that 
the Coulomb contribution to the elastic scattering rate is several times larger
than the kinetic-exchange contribution for typical chemical compositions.
The total scattering rate, averaged over the
majority heavy-hole Fermi surface, decreases 
with increasing density of itinerant
holes for a fixed Mn$^{2+}$ concentration.  
One important conclusion of this analysis is that
even in the heavily doped and compensated (Ga,Mn)As DMSs, the 
lifetime broadening of the quasiparticle 
($\hbar \Gamma$) is smaller than the valence band spin-orbit coupling strength 
($\Delta_{so}=341$~meV), 
which partially justifies the neglect of disorder\cite{jungwirthprl02} 
in evaluating some properties, {\it e.g.} the anomalous Hall conductivity.

In Fig.\ref{sigmaxx} we plot $\sigma_{xx}$, calculated numerically
using the six-band Kohn-Luttinger model and  Eqs.~(\ref{sigma})
and (\ref{gamma}),
for  a fully strained Mn$_{0.06}$Ga$_{0.94}$As sample grown on
a GaAs substrate and assuming compensation due to
As-antisites alone (i.e. no Mn-interstitials present). 
The substrate -- DMS lattice mismatch, $e_0\equiv(a_{sub}-a_{DMS})/a_{DMS}$, 
is between -0.002 and -0.003 in this case \cite{ohnosci98,gallagher}. 
(Note that $a_{sub}$ and $a_{DMS}$ are the
lattice constants of a fully relaxed substrate and ferromagnetic
layer, respectively.)
The parameters of the six-band Kohn-Luttinger model and strain coefficients 
used in these calculations are given in \cite{Meyer}.
We also show in Fig.\ref{sigmaxx} separate contributions
from individual heavy- and light-hole bands and demonstrate
that in the ferromagnetic state the current is carried 
mostly by the majority-spin heavy-holes, a property that will be important 
for understanding the spin-injection properties of (III,Mn)V DMS ferromagnets.
The absolute conductivities predicted by our model are reasonably 
consistent with experiment \cite{potashnik0204250,gallagher0205517,baxter0202508}.   
For lower Mn concentrations ($x<4$\%) the theoretical conductivities become several times
larger than the measured values due, we believe, to some combination of inaccuracy
in our scattering amplitude estimates,
sources of disorder in current
experimental samples that we do not account for in the model, 
and especially at small $x$ coherent scattering effects that eventually lead to localization.

As mentioned above, strong spin-orbit coupling in the
semiconductor valence band leads to a variety of magneto-anisotropy
effects \cite{dietlprb01,abolfathprb01,sinova0204209}. 
For dc transport the in-plane conductivity along, {\it e.g.} the 
x-direction should change when the magnetization
is rotated by applying a magnetic field stronger than the 
sample's magneto-crystalline anisotropy
field. In Fig.~\ref{amr} we show the anisotropic magnetoresistance
coefficients,
$AMR_{ip}\equiv[\rho_{xx}(\hat \Omega || x)-\rho_{xx}(\hat \Omega || y)]/
\rho_{xx}(\hat \Omega || y)$ and
$AMR_{op}\equiv[\rho_{xx}(\hat \Omega || x)-\rho_{xx}(\hat \Omega || z)]/
\rho_{xx}(\hat \Omega || z)$, for orthogonal magnetization directions in the
plane of the thin ferromagnetic layer, and for one of the magnetization directions
along the growth direction, respectively.
The Mn fraction assuming As-antisite compensation alone is indicated
in the figure by $x_{1}$.
The total Mn fraction (including substitutional Mn$^{2+}$ atoms and
Mn-interstitial atoms) for compensation due to interstitial Mn 
alone is labeled by $x_{2}$.
The plots demonstrate that in (Ga,Mn)As/GaAs ferromagnets,
which have compressive strain ($e_0<0$), $AMR$ is negative for typical
chemical compositions, and $|AMR_{ip}|< |AMR_{op}|$. 
(Note that $AMR_{ip}=AMR_{op}$ in unstrained cubic DMSs.)
The main plot shows that for a fixed hole density ($p=0.4$~nm$^{-3}$) 
the magnitude of the $AMR$ decreases from $\sim10\%$ to $\sim1\%$ with
increasing Mn fraction. All those observations are consistent with available 
experimental data on (Ga,Mn)As DMS's \cite{baxter0202508,gallagher}.
A more detailed comparison to measurements by 
Gallagher {\em et al.} \cite{gallagher} shows
that theoretical data assuming As-antisite compensation only overestimate
the decrease of $AMR$ with increasing Mn fraction. 
Better quantitative agreement is obtained
for compensation from Mn-interstitials, which is consistent with the presumed dominance
of Mn-interstitial defects over As-antisite defects in the samples measured by
Gallagher {\em et al.} \cite{gallagher}.

In addition to this $AMR$ engineering through doping that has been confirmed
experimentally, our theory predicts a large sensitivity of the spontaneous
transport anisotropy to strain. While strain does not
play a significant role for $AMR_{ip}$, as one might expect from symmetry 
considerations, $AMR_{op}$ can change by more than 10\% over the range of
strains that can be achieved in the thin ferromagnetic layers
by a proper choice of the substrate \cite{dietlprb01}. 
For larger Mn concentrations, $AMR_{op}$
is predicted to become positive in samples with tensile strain, as
shown in the main plot of Fig.~\ref{amr}.

We acknowledge helpful discussions with David Baxter, Jacek Furdyna, Bryan Gallagher, 
Bruce McCombe, and Peter Schiffer.  The work was supported by
the Welch Foundation, the DOE under grant DE-FG03-02ER45958, 
the Grant Agency of the Czech Republic under grant 202/02/0912
under grant OC P5.10, and by NSF MRSEC DMR-0080054 (M.A.).

%
%
\begin{figure}
\epsfxsize=3.20in
\centerline{\epsffile{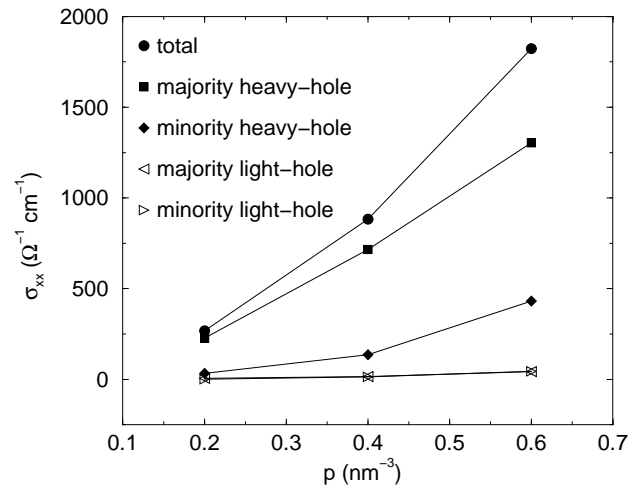}}
\caption{Conductivity for measuring current and magnetization
directed along the x-axis in the plane of the 
(Ga,Mn)As film as a function
of the total hole density. These results were obtained for a GaAs semiconductor host 
doped with 6\% Mn and with strain $e_0=-0.002$. No Mn-interstitial atoms are assumed to be present
in the ferromagnetic layer.
For typical hole densities the
current is carried mostly by the majority heavy-holes.}
\label{sigmaxx}
\end{figure}

%
%
%
\begin{figure}
\epsfxsize=3.40in
\centerline{\epsffile{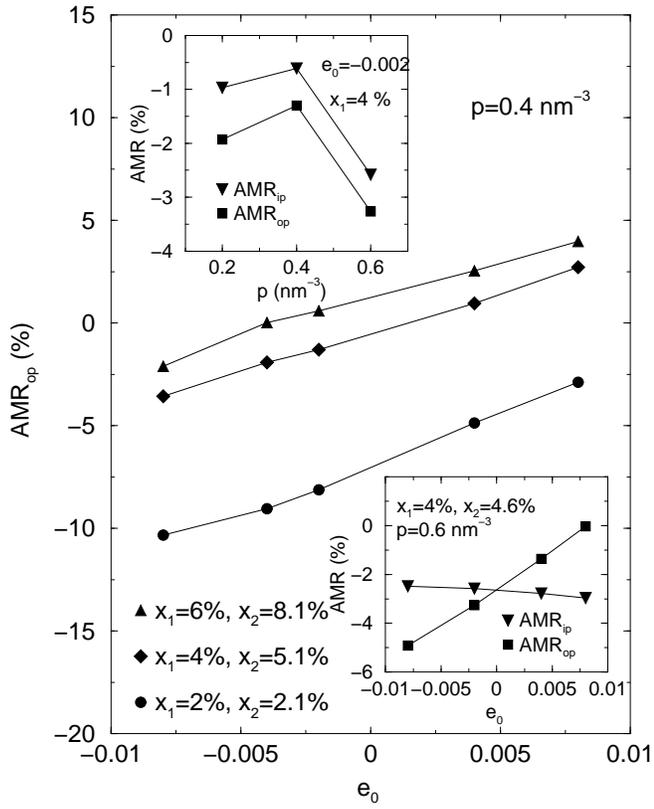}}
\caption{Anisotropic magnetoresistance coefficients
as a function of strain
(main plot and lower inset) and hole density (upper inset). Mn fractions
corresponding to compensation due to As-antisites alone
and Mn-interstitials alone are labeled as $x_{1}$ and $x_{2}$
respectively.}
\label{amr}
\end{figure}

\end{document}